\documentclass[prl,twocolumn,showpacs,superscriptaddress,preprintnumbers]{revtex4}
\usepackage{amssymb}
\usepackage{graphicx}
\usepackage{dcolumn}
\usepackage{bm}
\usepackage[bookmarks = true, pdfpagemode = None, pdfstartview = FitH,colorlinks = true,
citecolor = black, linkcolor = black, urlcolor = black]{hyperref}
\usepackage{url}

\newcommand{\ket}[1]{\left\vert{#1}\right\rangle}

\newcommand{\Phicox}{\Phi_{\text{co}}^x}
\newcommand{\Phiactx}{\Phi_{\text{act}}^x}

\begin{document}

\title{A Compound Josephson Junction Coupler for Flux Qubits With Minimal Crosstalk}
\author{R.~Harris}
\email{rharris@dwavesys.com}
\affiliation{D-Wave Systems Inc., 100-4401 Still Creek Dr., Burnaby, BC V5C 6G9, Canada}
\homepage{www.dwavesys.com}
\author{T.~Lanting}
\author{A.J.~Berkley}
\author{J.~Johansson}
\author{M.W.~Johnson}
\author{P.~Bunyk}
\author{E.~Ladizinsky}
\author{N.~Ladizinsky}
\author{T.~Oh}
\affiliation{D-Wave Systems Inc., 100-4401 Still Creek Dr., Burnaby, BC V5C 6G9, Canada}
\author{S.~Han}
\affiliation{Department of Physics and Astronomy, University of Kansas, Lawrence KS, USA}

\date{\today }

\begin{abstract}
An improved tunable coupling element for building networks of coupled rf-SQUID flux qubits has been experimentally demonstrated.  This new form of coupler, based upon the compound Josephson junction rf-SQUID,  provides a sign and magnitude tunable mutual inductance between qubits with minimal nonlinear crosstalk from the coupler tuning parameter into the qubits.  Quantitative agreement is shown between an effective one-dimensional model of the coupler's potential and measurements of the coupler persistent current and susceptibility.  
\end{abstract}

\pacs{85.25.Dq, 03.67.Lx}
\maketitle

The choice of architecture of a prototype solid state quantum information processor is primarily driven by the algorithm that the designer wishes to implement.  Within the field of superconducting quantum devices, at least two distinct architectures have arisen.  Gate model algorithms require qubits with long-lived excited states and dynamic couplings. Recent efforts have focused upon charge-like \cite{transmon} and phase \cite{martinis} qubits coupled to microwave resonators.  Adiabatic quantum algorithms \cite {Farhi,QA} require qubits whose groundstate encode binary variables and static couplings. One implementation involves a network of inductively coupled flux qubits \cite{architecture}.  The Hamiltonian for this architecture is that of a quantum Ising spin glass,
\begin{equation}
\label{eqn:Hprocessor}
{\cal H}_{\text{ISG}} = -\sum^N_{i=1}\frac{1}{2}\left[\epsilon_i\sigma_z^{(i)}+\Delta_i\sigma_x^{(i)}\right]+\sum_{i<j}J_{ij}\sigma_z^{(i)}\sigma_z^{(j)} \;\; ,
\end{equation}

\noindent where $\epsilon_i\equiv2\left|I_i^p\right|\Phi_i^x$ and $\Delta_i$ are the bias and tunneling energy of qubit $i$, respectively, and $J_{ij}\equiv M_{ij}\left|I_i^p\right|\left|I_j^p\right|$ is the coupling energy between qubits $i$ and $j$.  Here, $\left|I_i^p\right|$ represents the magnitude of the qubit persistent current, $\Phi_i^x$ is an external flux bias and $M_{ij}$ is a mutual inductance.  
A programmable processor would require in-situ tunable $\Phi_i^x$ and $M_{ij}$.  Inductive coupling could also be useful in other quantum computation schemes in which the flux qubit's persistent current basis is nearly concurrent with the computation basis \cite{fluxbasis}.  More involved parametric coupling schemes are needed if the flux qubits are biased to their optimal points where the energy and persistent current bases are orthogonal \cite{energybasiscoupling}.  

The authors of Ref.~\cite{AMvdB} proposed the use of an rf-SQUID to implement tunable $M_{ij}$.  Experiments on systems of coupled flux qubits verified that such couplers did perform as anticipated \cite{AMvdBCoupler}.  However, additional work not reported in the literature revealed two serious deficiencies:  First, the tuning mechanism involves threading flux through the rf-SQUID loop, thus inducing a large persistent current $I_p$ that, in turn, biases the qubits.  This  is a significant problem if the qubit biases need to be controlled to high precision atop what can be a very large nonlinear crosstalk imparted by the coupler.  Second, $M_{ij}=0$ can only be achieved if $\beta\equiv 2\pi L I_c/\Phi_0<1$, where $L$ and $I_c$ are the rf-SQUID inductance and critical current, respectively.  On the other hand, in order to achieve appreciable non-zero coupling it proved necessary to design devices with $\beta\gtrsim 0.9$.  Such devices were acutely sensitive to fabrication variations, where higher than expected $I_c$ could make $M_{ij}=0$ unattainable.  Thus the rf-SQUID coupler proved troublesome in practice.  Note that the dc-SQUID \cite{AMvdB,DCSQUIDCoupler} and the unipolar \cite{UnipolarCoupler} couplers suffer from similar deficiencies.  The challenge was then to design a tunable $M_{ij}$ that invokes minimal $I_p$ and is robust against fabrication variations.

\begin{figure}[tbp]
\includegraphics[width=2.5in]{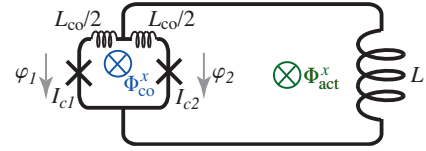}
\caption{(Color online)  Schematic of a CJJ rf-SQUID.}
\label{fig:CJJRFS}
\end{figure}

Our efforts to identify a satisfactory coupler design led us to consider the compound Josephson junction (CJJ) rf-SQUID, as depicted in Fig.~\ref{fig:CJJRFS}.  The CJJ rf-SQUID comprises a superconducting loop of inductance $L$ that is interrupted by a smaller loop of inductance $L_{\text{co}}$ containing two Josephson junctions with critical currents $I_{c1}$ and $I_{c2}$.  Devices with this general topology can take on a number of guises:  The CJJ rf-SQUID can be used as a qubit when designed with relatively low $I_c=I_{c1}+I_{c2}\lesssim 3\,\mu$A, $L$ chosen such that $\beta\lesssim 2$, low net capacitance across the junctions $C$ and biased such that its potential energy is bistable \cite{CJJqubit,synchronization}.  The CJJ rf-SQUID can also be used as a latching readout when designed with a substantial $I_c\gtrsim 10\,\mu$A, $L$ chosen such that $\beta\gtrsim 1.2$, large $C$ and having its potential swept from being monostable to bistable \cite{qfp}.  We focus herein upon a CJJ rf-SQUID designed with a modest $I_c$, $L$ chosen such that $\beta\lesssim 1.2$ and operated with a monostable potential.  In Fig.~\ref{fig:CJJRFS} the closed loops are subjected to external flux biases $\Phiactx\equiv\Phi_0\varphi_{\text{act}}^x/2\pi$ and $\Phicox\equiv\Phi_0\varphi_{\text{co}}^x/2\pi$.  The symbols have been chosen to indicate that $\Phiactx$ is an actuator for perturbing the device and $\Phicox$ represents the control signal.  Let the phase across the junctions be $\varphi_1$ and $\varphi_2$.  The Hamiltonian for this device can be written as 
\begin{equation}
\label{eqn:2JHphase}
{\cal H}=\sum_{i=1}^2\left[\frac{Q_i^2}{2C_i}-E_{Ji}\cos(\varphi_i)\right]+\sum_{n}U_n\frac{\left(\varphi_n-\varphi_n^x\right)^2}{2}
\end{equation}

\noindent where $C_{i}$ and $E_{Ji}=I_{ci}\Phi_0/2\pi$ represent the capacitance and Josephson energy of junction $i$, respectively, and $[\Phi_0\varphi_i/2\pi,Q_j]=i\hbar\delta_{ij}$.  The inductive terms originate from the two closed loops with $n\in\left\{\text{co},{\text{act}}\right\}$, $L_{\text{act}}\equiv L+L_{\text{co}}/4$ and $U_n\equiv(\Phi_0/2\pi)^2/L_n$.  The actuator and control loop phases are defined as $\varphi_{\text{act}}\equiv\left(\varphi_1+\varphi_2\right)/2$ and $\varphi_{\text{co}}\equiv\varphi_1-\varphi_2$, respectively.  Hamiltonian (\ref{eqn:2JHphase}) can be reduced to an effective 1-dimensional system if $L_{\text{act}}\gg L_{\text{co}}$ because the plasma energy of the control loop will then be much higher than that of the actuator loop.  Setting $\varphi_{\text{co}}=\varphi_{\text{co}}^x$ and combining the Josephson terms,
\begin{equation}
\label{eqn:2JHeff}
{\cal H}\approx \frac{Q_{\text{act}}^2}{2C_p}+V(\varphi_{\text{act}})
\end{equation}
\vspace{-15pt}
\begin{displaymath}
V(\varphi_{\text{act}})=U_{\text{act}}\Big\{\frac{\left(\varphi_{\text{act}}-\varphi_{\text{act}}^x\right)^2}{2}-\beta_{\text{eff}}\cos\left(\varphi_{\text{act}}-\varphi_{\text{act}}^0\right)\Big\}
\end{displaymath}
\vspace{-15pt}
\begin{displaymath}
\beta_{\text{eff}}\equiv\frac{2\pi L_{\text{act}}I_{c+}}{\Phi_0}\cos\left(\frac{\varphi_{\text{co}}^x}{2}\right)\sqrt{1+\left[\frac{I_{c-}}{I_{c+}}\tan\left(\frac{\varphi_{\text{co}}^x}{2}\right)\right]^2}
\end{displaymath}
\vspace{-15pt}
\begin{displaymath}
\varphi_{\text{act}}^0\equiv -\arctan\left[\frac{I_{c-}}{I_{c+}}\tan{\frac{\varphi_{\text{co}}^x}{2}}\right]
\end{displaymath}

\noindent where $I_{c\pm}\equiv I_{c1}\pm I_{c2}$ and $C_p=C_1+C_2$.  Hamiltonian (\ref{eqn:2JHeff}) is homologous to that of an rf-SQUID whose single junction possesses a critical current that is a function of $\varphi_{\text{co}}^x$ and whose phase has been shifted by $\varphi_{\text{act}}^0$.

Let the device described by Eq.~(\ref{eqn:2JHeff}) be connected to two qubits via mutual inductances $M_{\text{co},1}$ and $M_{\text{co},2}$.  The mutual inductance between the qubits will be
\begin{equation}
\label{eqn:mij}
M_{\text{eff}}=M_{\text{co},1}M_{\text{co},2}\chi^{(1)}
\end{equation}

\noindent where $\chi^{(1)}\equiv\partial I_{\text{act}}^p/\partial\Phiactx$ represents the first order (linear) susceptibility of the coupler \cite{AMvdB} and the persistent current flowing about the coupler actuator loop is
\begin{equation}
\label{eqn:Ip}
I_{\text{act}}^p\equiv\frac{\beta_{\text{eff}}}{2\pi L_{\text{act}}/\Phi_0}\sin\left(\varphi_{\text{act}}-\varphi_{\text{act}}^0\right) \;\; .
\end{equation}

\noindent If $V(\varphi_{\text{act}})$ is monostable and the first excited state can be neglected, then one can replace the operator $\varphi_{\text{act}}$ by the value for which $V$ is a minimum ($dV/d\varphi_{\text{act}}=0$):
\begin{equation}
\label{eqn:fluxquantization}
\varphi_{\text{act}}-\varphi_{\text{act}}^x+\beta_{\text{eff}}\sin\left(\varphi_{\text{act}}-\varphi_{\text{act}}^0\right)=0 ,
\end{equation}

\noindent which can be solved for $\varphi_{\text{act}}$ given arbitrary $\varphi_{\text{act}}^x$, thus yielding $I_{\text{act}}^p(\Phiactx,\Phicox)$.  Differentiating Eqs.~\ref{eqn:Ip} and \ref{eqn:fluxquantization}
with respect to $\Phiactx$ then yields $\chi^{(1)}$:
\begin{equation}
\label{eqn:chi}
\chi^{(1)}\equiv\frac{\partial I_{\text{act}}^p}{\partial\Phiactx}=\frac{1}{L_{\text{act}}}\frac{\beta_{\text{eff}}\cos\left(\varphi_{\text{act}}-\varphi_{\text{act}}^{\text{0}}\right)}{1+\beta_{\text{eff}}\cos\left(\varphi_{\text{act}}-\varphi_{\text{act}}^{\text{0}}\right)}
\end{equation}

\noindent Equation (\ref{eqn:chi}) is similar to Eq.~(10) of Ref.~\cite{AMvdB}, albeit $\beta_{\text{eff}}$ is a function of $\varphi_{\text{co}}^x$ and junction asymmetry results in a $\varphi_{\text{co}}^x$-dependent phase shift in the cosine terms.

While rf-SQUID and CJJ rf-SQUID couplers possess similar expressions for $\chi^{(1)}$, the latter holds two advantages:  First, the CJJ coupler can be operated with $\Phiactx=0$ and tuned via $\Phicox$.  If $I_{c-}/I_{c+}\ll 1$, then $\varphi_{\text{act}}^0\ll 1$ and Eq.~(\ref{eqn:fluxquantization}) yields $\varphi_{\text{act}}\approx 0$.  Equation (\ref{eqn:Ip}) then predicts that $I_{\text{act}}^p\approx 0$.  Thus the CJJ coupler need not invoke large persistent currents (on the order of $I_{c+}$) when being tuned.  Second, the CJJ coupler is usable over the range of $\Phicox$ for which $-\text{min}\left[1, \beta_{\text{eff}}(0)\right]\lesssim\beta_{\text{eff}}(\varphi_{\text{co}}^x)\leq\beta_{\text{eff}}(0)$ when $\Phiactx=0$, where the lower bound has been imposed by the condition that $V(\varphi_{\text{act}})$ be monostable.  Thus the utility of the CJJ coupler is not compromised if $\beta_{\text{eff}}(0)>1$.  As such, this device is robust against fabrication variations.

To test the CJJ rf-SQUID coupler, we fabricated a circuit containing 8 CJJ rf-SQUID flux qubits \cite{CJJqubit,synchronization}, each inductively coupled to its own hysteric dc-SQUID readout \cite{readout}, and connected by a network of 16 CJJ rf-SQUID couplers.  The chip was fabricated from an oxidized Si wafer with Nb/Al/Al$_2$O$_3$/Nb trilayer junctions, four Nb wiring layers capped with SiN and separated by planarized PECVD SiO$_{2}$.  The chip was mounted to the mixing chamber of a dilution refrigerator regulated at $T=40\,$mK inside a Sn superconducting magnetic shield with a residual field in the vicinity of the chip $\lesssim 1\,$nT.  External current biases were provided by room temperature current DACs whose outputs were low pass filtered with $f_c\approx 5\,$MHz using a combination of lumped element and copper powder filters secured to the mixing chamber.

\begin{figure}[tbp]
\includegraphics[width=3.2in]{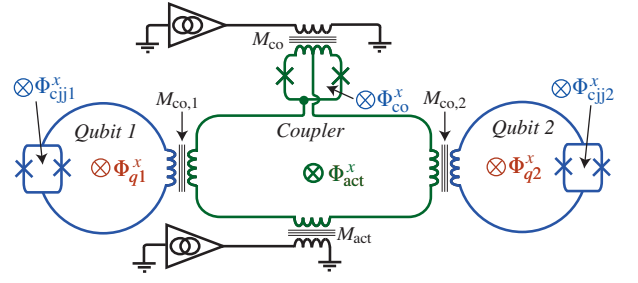}
\caption{(Color online)  Schematic of a CJJ rf-SQUID coupler interacting with two CJJ rf-SQUID qubits.}
\label{fig:Coupler}
\end{figure}

A schematic of a coupler and a pair of qubits is depicted in Fig.~\ref{fig:Coupler}.  The coupler is controlled via bias currents that are coupled to the device through mutual inductances $M_{\text{co}}$ and $M_{\text{act}}$, respectively.  These give rise to the fluxes $\Phicox$ and $\Phiactx$. The qubits are controlled via fluxes $\Phi_{\text{cjj}\alpha}^x$ and $\Phi_{q\alpha}^x$ ($\alpha=1,2$) as described in Ref.~\cite{synchronization}.  The qubits interact with the coupler via mutual inductances $M_{\text{co},\alpha}$.  For brevity, we present results from a single coupler in this paper and note that $M_{\text{eff}}(\Phicox)$ was identical to $\lesssim 5\%$ for all 16 couplers on this chip.  For the particular coupler described herein, the relevant qubit critical currents were $I_{q\alpha}^c=3.25\pm 0.01\,\mu$A and qubit inductances were $L_{q1(2)}=290 (308)\pm 5\,$pH when $\Phicox=0$.

\begin{figure}[tbp]
\includegraphics[width=3.2in]{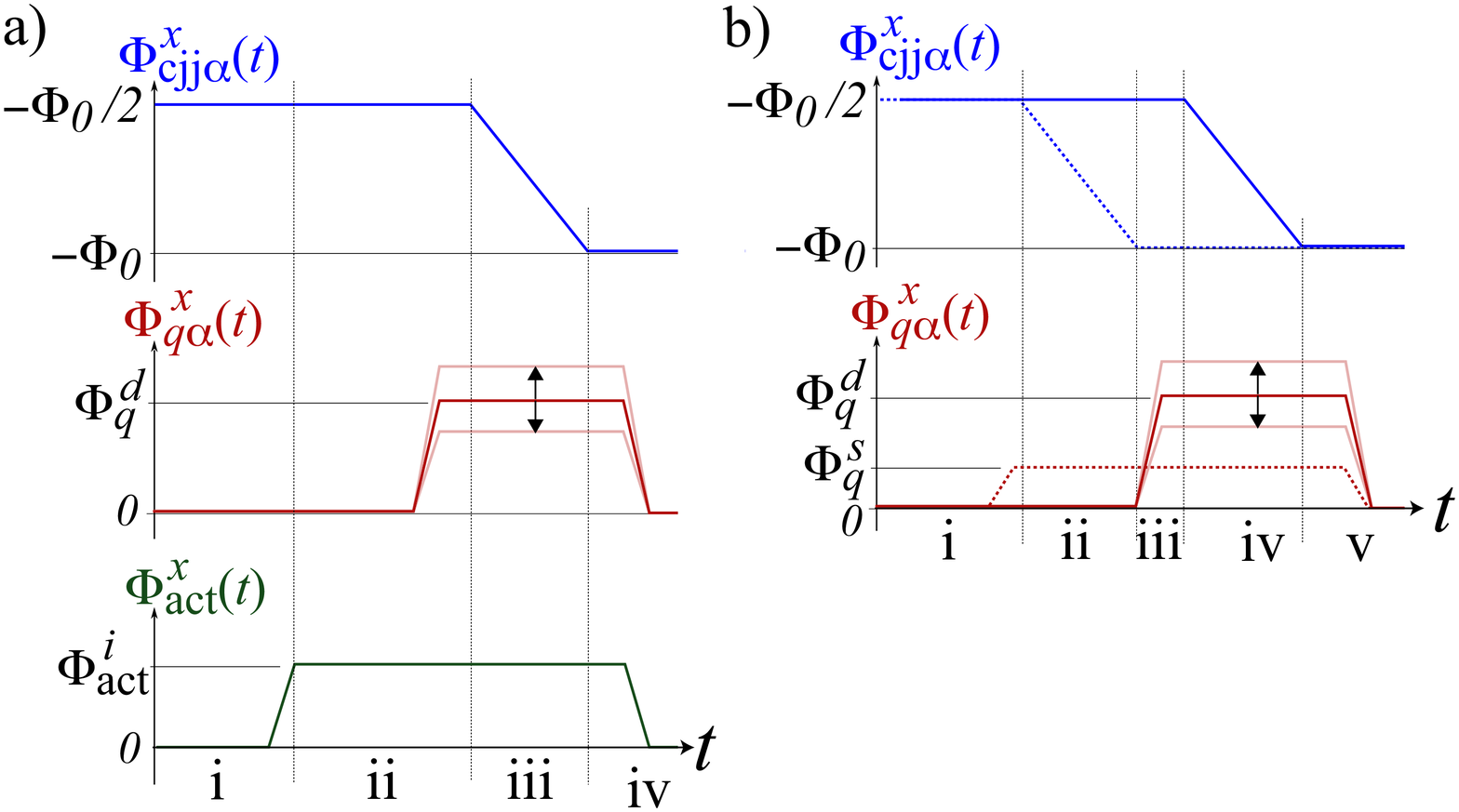}
\caption{(Color online) a) Single qubit measurement waveforms. b) Sequential annealing waveforms.  Source (detector) qubit waveforms denoted as dashed (solid) lines.}
\label{fig:qubitwaveforms}
\end{figure}

The flux waveforms used to obtain $M_{\text{act}}$ are depicted in Fig.~\ref{fig:qubitwaveforms}a.  In this case, $\Phicox$ was held constant while the detector qubit ($\alpha=d$) was annealed in the presence of a pulse on $\Phiactx(t)$ of amplitude $\Phi_{\text{act}}^i$ and a pulse on $\Phi_{q\alpha}^x(t)$ of amplitude $\Phi_q^d$.  The sequence involves initializing the qubit in a monostable potential with no net flux biases (i), setting $\Phiactx$ and $\Phi_{q\alpha}^x$ (ii), raising the detector qubit's tunnel barrier to maximum height (iii) and then returning $\Phiactx$ and $\Phi_{q\alpha}^x$ to zero prior to reading the state of the detector qubit (iv).  The result of this process is that the state of the detector qubit becomes trapped in one of its two counter-circulating persistent current states, denoted as $\ket{0}$ and $\ket{1}$.  Repeating this sequence to gather statistics then yielded the probability of finding the detector in $\ket{0}$, $P_0$.  Using software feedback, we adjusted $\Phi_q^d$ to track the displacement of the detector's degeneracy point $\Phi_{q\alpha}^0$, defined as the bias for which $P_0=1/2$, to within $\pm 0.02\,$m$\Phi_0$.  We have defined $\Phi_{q\alpha}^0\equiv 0$ with respect to the degeneracy point obtained with $\Phicox=0$ and $\Phi_{\text{act}}^x=0$.  Mapping $\Phi_{q\alpha}^0$ versus the current bias driving $\Phi_{\text{act}}^i$ yielded a modulation with period $\Delta I_{\text{act}}$, from which we obtained $M_{\text{act}}=\Phi_0/\Delta I_{\text{act}}=1.77\pm 0.01\,$pH.

To obtain $M_{\text{co}}$ we again used the flux waveform pattern depicted in Fig.~\ref{fig:qubitwaveforms}a but with $\Phi_{\text{act}}^i$ toggled between $\pm 5\,$m$\Phi_0$.  Taking the difference in $\Phi_{q,\alpha}^0$ between the two polarizations, we tracked the amount of coupled flux $X_1^{\alpha}\equiv 2M_{\text{co},\alpha}\chi^{(1)}(\Phicox)\Phi_{\text{act}}^i$ versus the bias driving $\Phicox$.  The results yielded a period $\Delta I_{\text{co}}$, from which we obtained $M_{\text{co}}=\Phi_0/\Delta I_{\text{co}}=3.43\pm 0.03\,$pH.

\begin{figure}[bp]
\includegraphics[width=3.2in]{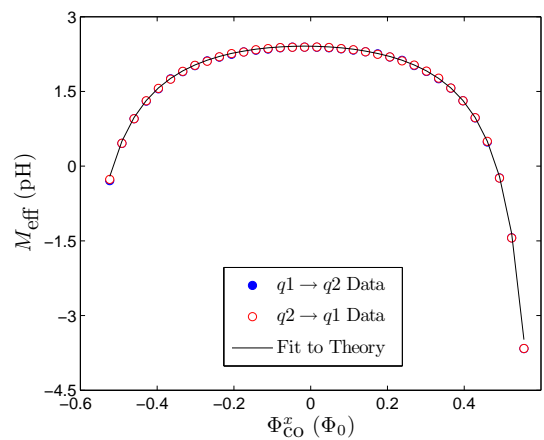}
\caption{(Color online) CJJ rf-SQUID coupler effective mutual inductance versus control flux. Solid (hollow) points correspond to $\alpha=1(2)$ acting as flux source and $\alpha=2(1)$ as flux detector.  Solid curve is from a simultaneous best fit of these data to Eqs.~(\ref{eqn:mij}) and (\ref{eqn:chi}) and those in Fig.~\ref{fig:DegPtMotion}a to Eq.~(\ref{eqn:Ip}).}
\label{fig:chi}
\end{figure}

With the coupler biases calibrated, we proceeded with measuring $M_{\text{eff}}(\Phicox)$.  To do so, we used the 2-qubit flux bias sequence depicted in Fig.~\ref{fig:qubitwaveforms}b in which one qubit served as a flux source ($\alpha=s$) and the other acted as a flux detector ($\alpha=d$).  This process, referred to as sequential annealing, involved initializing both qubits in monostable potentials with no net flux biases (i), setting $\Phi_{qs}^x=\Phi_q^s=\pm 5\,$m$\Phi_0$ and raising $\Phi_{\text{cjj},s}^x$ to trap the source qubit in either $\ket{0}$ or $\ket{1}$ (ii), using software feedback to adjust $\Phi_q^d$ (iii) and raising the detector qubit's tunnel barrier to trap its state (iv).  Finally, both $\Phi_{q\alpha}^x$ were returned to zero prior to reading the state of the detector qubit (v).  The relative change in $\Phi_{qd}^0$ between the two polarizations of the source qubit then yielded the flux $X_2^{(s)}\equiv 2M_{\text{eff}}(\Phicox)\left|I_{s}^p\right|$.  For each qubit, $\left|I_{\alpha}^p\right|$ could be directly inferred from measurements obtained with its readout.  Measurements of $M_{\text{eff}}(\Phicox)$ are shown in Fig.~\ref{fig:chi}.  Using any three of $(X_1^{(1)},X_1^{(2)},X_2^{(1)},X_2^{(2)})$, one could also solve for $M_{\text{co},1}=M_{\text{co},2}=17.5\pm 0.2\,$pH and $\chi^{(1)}(\Phicox)$.  

\begin{figure}[tbp]
\includegraphics[width=3.05in]{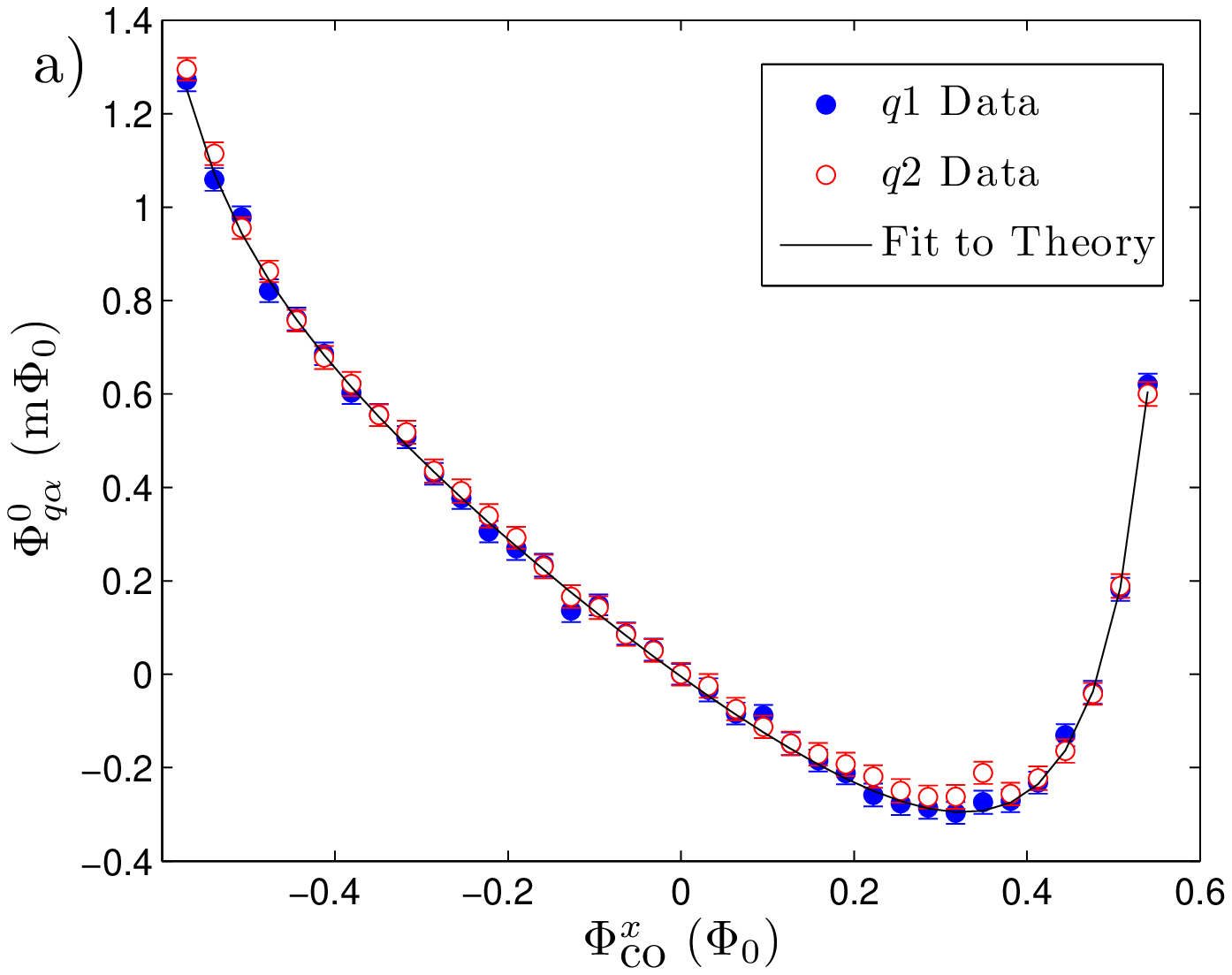} \\
\includegraphics[width=3.05in]{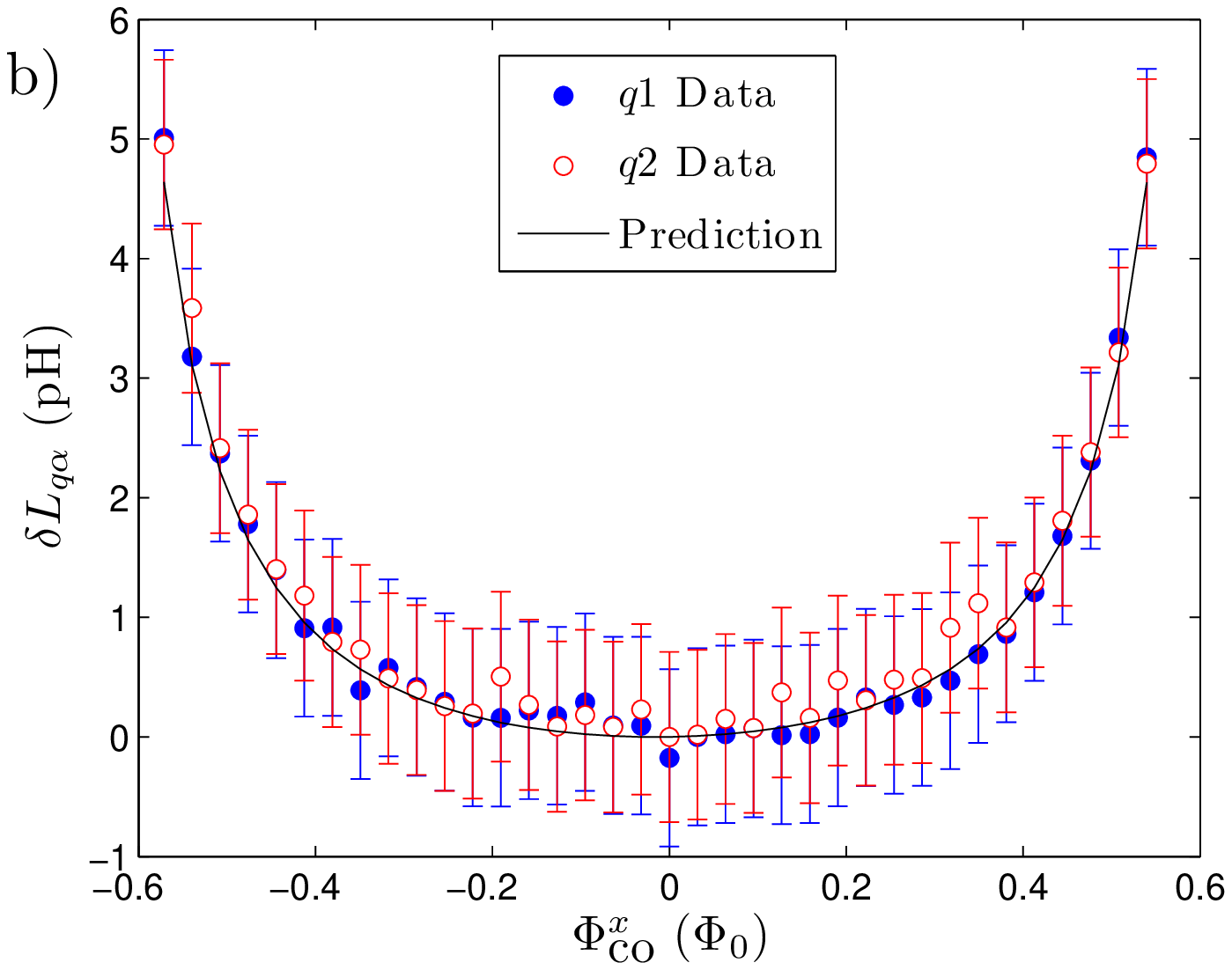}
\caption{(Color online) Deviation of qubit parameters as a function of coupler control flux: a) Qubit degeneracy point.  Solid curve is from a simultaneous best fit of these data to Eq.~\ref{eqn:Ip} and those in Fig.~\ref{fig:chi} to Eq.~(\ref{eqn:chi}).  b) Qubit inductance.  Solid curve is prediction using the best fit parameters.} 
\label{fig:DegPtMotion}
\end{figure}

A key motivation for developing the CJJ coupler was to minimize the impact of coupler settings upon qubit properties.  
Plots of the relative change in qubit degeneracy point $\Phi_{q\alpha}^0=M_{\text{co},\alpha}I_{\text{act}}^p$ versus $\Phicox$ are shown in Fig.~\ref{fig:DegPtMotion}a.  The qubit inductance $L_{q\alpha}$ will also be a function of $\Phicox$.  Let the change in inductance be defined as
\begin{equation}
\label{eqn:deltaL}
\delta L_{q\alpha}(\Phicox)=M_{\text{co},\alpha}^2\left[\chi^{(1)}(\Phicox)-\chi^{(1)}(0)\right]\;\; .
\end{equation}

\noindent  We have quantified this effect by measuring $\left|I_{\alpha}^p\right|$ versus $\Phicox$.  For $\Phi_{\text{cjj},\alpha}^x=-\Phi_0$ and $\Phi_{q\alpha}^x=0$ one can use an approximation, similar to that used to obtain Eq.~(\ref{eqn:Ip}), to write a pair of expressions for $\left|I_{\alpha}^p\right|$:
\begin{eqnarray}
\label{eqn:iqp}
\left|I_{\alpha}^p\right| = I_{q\alpha}^c\left|\sin\varphi_q\right| \\
\varphi_q-\frac{2\pi L_{q\alpha}I_{q\alpha}^c}{\Phi_0}\sin\varphi_q=0 \nonumber
\end{eqnarray}

\noindent Given the calibrated values of $I_{q\alpha}^c$ for each qubit, it was then possible to extract $L_{q\alpha}(\Phicox)$ from $\left|I_{\alpha}^p(\Phicox)\right|$.  The resultant $\delta L_{q\alpha}= L_{q\alpha}(\Phicox)-L_{q\alpha}(0)$ are shown in Fig.~\ref{fig:DegPtMotion}b.  

The data shown in Figs.~\ref{fig:chi} and \ref{fig:DegPtMotion}a have been simultaneously fit to $M_{\text{co},1}M_{\text{co},2}\chi^{(1)}$ [Eqs.~(\ref{eqn:mij}) and (\ref{eqn:chi})] and $M_{co,\alpha}I_{\text{act}}^p$ [Eq.~(\ref{eqn:Ip})], respectively.  In order to yield a high quality fit, we allowed for small flux offsets in both coupler loops and a small linear crosstalk from the control bias to the actuator loop: $\Phicox\rightarrow\Phicox-\Phi_{\text{co}}^0$ and $\Phiactx\rightarrow\gamma\Phicox-\Phi_{\text{act}}^0$.  The best fit was obtained with $I_{c+}=8.4\pm 0.3\,\mu$A, $I_{c-}/I_{c+}=(6.7\pm 0.9)\times 10^{-3}$, $L_{\text{act}}=88\pm 2\,$pH,  $\Phi_{\text{co}}^0=16\pm 1\,$m$\Phi_0$,  $\Phi_{\text{act}}^0=3\pm 1\,$m$\Phi_0$ and $\gamma=(6\pm 1)\times 10^{-3}$.  Note that direct coupling between qubits would add a positive constant to $M_{\text{eff}}$ [Eq.~(\ref{eqn:mij})].  A key consequence would be that the difference between the two values of $\Phicox$ where $M_{\text{eff}}=0$ would be {\it greater} than $\Phi_0$, as predicted by Eq.~(\ref{eqn:chi}).  The best fit curve in Fig.~\ref{fig:chi} suggests that $\Delta\Phicox|_{M_{\text{eff}}=0}\approx 0.998\,\Phi_0$, therefore the direct coupling appears to be negligible.

The solid curve in Fig.~\ref{fig:DegPtMotion}b represents the predicted $\delta L_{q\alpha}$ using the best fit parameters.  Given the agreement between theory and experiment, one can conclude that Eqs.~(\ref{eqn:2JHeff})$\rightarrow$(\ref{eqn:chi}) correctly model the physics of this device.  Note that over the bias range 
shown that the qubit degeneracy points shift by $\sim 2\,$m$\Phi_0\ll M_{\text{co},\alpha}I_{c+}\sim 70\,$m$\Phi_0$.  Consequently, one can conclude that the nonlinear crosstalk from coupler to qubit is substantially less than that encountered while tuning a comparable single junction rf-SQUID coupler \cite{AMvdB}.  According to Eq.~\ref{eqn:Ip}, this undesirable effect could be reduced to negligible levels by improvements in fabrication uniformity (smaller $I_{c-}/I_{c+}$).  Achieving lower $\gamma$ in future designs will also help realize further reductions in nonlinear flux offsets.

Changes in $L_{q\alpha}$ are of consequence if the properties of multiple qubits need to be synchronized to high precision \cite{synchronization}.  Custom tuned qubit CJJ flux offsets provide one means of mitigating this undesirable effect \cite{synchronization}.  Alternate qubit designs which contain an in-situ tunable inductance for ballast constitute a second solution.

{\it Conclusions:}  A compound Josephson junction rf-SQUID coupler suitable for building networks of coupled flux qubits has been demonstrated.  This coupler provides both sign and magnitude tunable mutual inductance in a manner that invokes minimal nonlinear crosstalk from the coupler tuning parameter to the qubits.  Furthermore, this crosstalk can be reduced to negligible levels with improved fabrication uniformity and subtle improvements in device layout.  Modulation of the qubit inductance via changes in the coupler settings has been characterized and shown to be predictable using an effective one-dimensional model of the coupler potential.

We thank J.~Hilton, G.~Rose, P.~Spear, A.~Tcaciuc, F.~Cioata, E.~Chapple, C.~Rich, C.~Enderud, B.~Wilson, M.~Thom, S.~Uchaikin, M.~Amin, F.~Brito, D.~Averin, A.~Kleinsasser and G.~Kerber. S.Han was supported in part by NSF Grant No. DMR-0325551.

\end{document}